\documentclass[aps,groupedaddress]{revtex4}
\pdfoutput=1
\usepackage{graphicx}
\usepackage{amsmath}
\usepackage{setspace}

\begin{document}

\title{A Tale of Two Tilings}

\author{Sharon C. Glotzer$^{1,2}$}
\author{Aaron S. Keys$^{1}$}
\affiliation{$^1$Department of Chemical Engineering and \\ $^2$Department of Materials Science Engineering \\ University of Michigan, Ann Arbor, Michigan 48109-2136}

\date{\today}

\begin{abstract}
\textbf{What do you get when you cross a crystal with a quasicrystal?  The surprising answer stretches from Fibonacci to Kepler, who nearly 400 years ago showed how the ancient tiles of Archimedes form periodic patterns.}
\end{abstract}

\maketitle

\setstretch{1.5}

Quasicrystals are mosaic-like arrangements of atoms that have symmetries once thought to be impossible for crystals to adopt$^1$. Primarily observed in certain metal alloys, these unusual structures whose patterns adorn medieval Islamic mosques and palaces$^2$ are stronger and less deformable than analogous regular crystals, and have unusual frictional, catalytic and optical properties. Several applications for quasicrystals have been proposed Ñ for example, some could be used as materials for circuits that are based on the flow of photons, rather than electrons$^3$. But for this application to be realized, the atomic dimensions of a quasicrystal must first be scaled up nearly a thousand-fold. In reference 4, Mikhael et al. describe quasicrystals at just such a scale, made from microscopic, plastic beads. To their surprise, they also discovered a new kind of structure: a rare type of 1d quasicrystal that can be considered a cross between a 2d quasicrystal and a regular crystal. 

Mikhael et al. grow single layers of colloidal beads, or particles, on a templated surface designed to attract particles and arrange them into pentagons Ñ the primary motif of a quasicrystal with ten-fold (decagonal) symmetry. They do this by arranging five laser beams to form an interference pattern that confers decagonal symmetry to the surfaceÕs potential, which interacts with the particles$^5$. By tuning the strength of the surface potential using the lasers, the team controls the formation of the growing structures: regular crystals form when particleÐparticle interactions dominate, and quasicrystals form when particleÐsurface interactions dominate. The resulting quasicrystals exhibit interlocking rings of ten particles surrounding a central particle (see Fig. 1c of reference 4). 

Quasicrystals are often considered to be intermediate between glasses (amorphous solids) and crystals$^6$. But can a structure be intermediate between a crystal and a quasicrystal?  Conventional thinking says no Ñ long range ordering must be either periodic (crystalline) or aperiodic (quasicrystalline) with little room in between. But Mikhael et al. find that, when the particleÐparticle and particleÐsurface interactions in their system are similar in strength, an intermediate phase forms that combines elements of both crystalline and quasicrystalline ordering. In fact, the particles assemble into something closely resembling an Archimedean tiling pattern.

Archimedean tilings are periodic arrangements of regular polygons laid edge-to-edge in a plane.  Their defining feature is that only one kind of vertex must exist Ñ that is, where the corners of the polygons meet at a point, any given corner must always meet the same combination of corners from other polygons. Archimedean tilings have been used in art and architecture since antiquity, but it was the astronomer Johannes Kepler who first classified them in his book, Harmonices Mundi, in 1619. Kepler showed that there are eleven different kinds of tilings, eight of which contain more than one type of regular polygon.   One type of tiling consists entirely of equilateral triangles, and is denoted ($3^6$) to indicate that six triangles meet at each vertex.  This structure describes the crystal that Mikhael, et al. observe when particle-particle interactions dominate (see Fig. 2a of reference 4).  Another Archimedean tiling denoted ($3^3$,$4^2$) consists of alternating rows of squares and triangles (see Fig. 3a of reference 4). 

\begin{figure*}
\includegraphics[width=0.7\columnwidth]{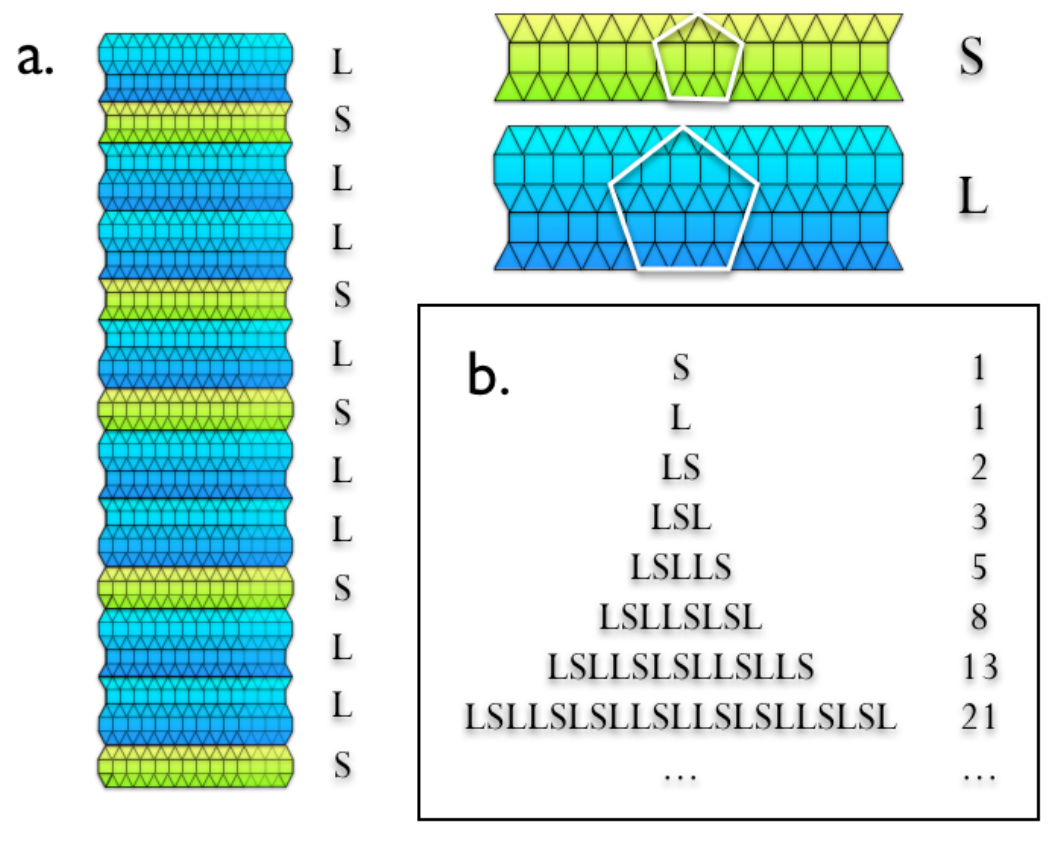}
\caption{Archimedean tiling with aperiodic ÒdefectsÓ and its relation to a Fibonacci chain. (a) Ideal schematic of the intermediate structure found by Mikhael, et al. Colloidal particles sit at the vertices of the tiles. (b) Fibonacci chain formed by applying the substitution rules L$\rightarrow$LS, S$\rightarrow$L at each step. The sequence with 13 elements describes the arrangement of Òunit-cellsÓ in the structure shown in (a).}
\end{figure*}

Mikhael and colleaguesÕ new arrangement of particles is similar to the ($3^3$,$4^2$) arrangement, with some ($3^6$) vertex configurations added in a peculiar way. The particles form alternating rows of squares and triangles, interrupted intermittently by ÔdefectsÕ Ñ additional rows of triangles (see Fig. 3b of reference 4). The particles still align locally with the decagonal, quasicrystalline template, but a mismatch between the periodic tiling and the aperiodic substrate arises over longer distances. This is where the defects come in Ñ the extra rows of triangles correct the mismatches.  

The defects result in two distinct ``unit cells.''  These cells are stacked in a quasiperiodic pattern known as a ÒFibonacci chainÓ (see Fig. 1). Named for the famous Italian mathematician Leonardo Fibonacci, this pattern is often found in nature, and describes the structure of one-dimensional quasicrystals$^1$.  In Mikhael and colleaguesÕ system, the Fibonacci chain determines the sequence of long (L) and short (S) cells.  The heights of the cells correspond to the heights of large and small pentagons conferred by the field.  Although it is not discussed by the authors, the Fibonacci chain is self-similar, and thus the structure is equivalently described by simpler unit cells consisting of single rows of squares (S) and triangles (L).
When grown on an icosahedral quasicrystalline surface, certain copper alloys also adopt a curious phase in which the atoms have a Fibonacci spacing$^7$. The exact structure of the phase has not yet been identified, but its diffraction pattern is identical to that of Mikhael et alÕs Archimedean-like arrangement of particles. If the two phases are indeed identical, it would demonstrate the universality of the underlying physics that controls the templated growth of these unusual structures and further extend the growing use of colloids as ÒminimalÓ models of atoms for studying self-assembly$^8$ and other physical processes. 

Interestingly, Archimedean tilings also form from macromolecules that consist of three chemically distinct polymers, covalently bonded together at one end to form a three armed ÔstarÕ$^9$. Under certain conditions, these systems spontaneously form cylinders that have a cross-section corresponding to one of four Archimedean tilings. Two of these structures have useful optical properties, and, like quasicrystals, hold promise for photonic applications$^9$. 

It is not clear whether Archimedean-like tilings have a general role as intermediates between periodic and aperiodic structures. Such intermediates must be able to locally align with both the corresponding quasicrystal and crystal structures, and be able to incorporate aperiodically-arranged defects.  The ability to mix and match vertex configurations may give Archimedean-tiling motifs a unique flexibility that makes them prone to forming aperiodic arrangements. For example, the dodecagonal quasicrystal$^10$, which exhibits 12-fold, rather than 10-fold, rotational symmetry, is made up of three different Archimedean vertex configurations. 

It is important to note that we should not think of Mikhael and colleaguesÕ structure as a flawed Archimedean tiling.  The underlying structure is a perfect Fibonacci chain whose elements are decorated with infinite rows of Archimedean tiles. From this perspective, it is a unique kind of one-dimensional quasicrystal, periodic in one dimension, but quasiperiodic in the other.  This is what you get when you cross a crystal with a quasicrystal Ñ a beguiling new tiling built upon iconic mathematical foundations. 

\textit{Sharon C. Glotzer and Aaron S. Keys are in the Department of Chemical Engineering, University of Michigan, Ann Arbor, Michigan 48109-2136, USA. \\
e-mail: sglotzer@umich.edu     
}

\begin{enumerate}
\setstretch{1.0}
\item Janot, C. Quasicrystals: A Primer (Oxford Univ. Press, 1997).
\item Lu, P. J. and Steinhardt, P. J. Science 315, 1106-1110 (2007).
\item Man, W., Megens, M., Steinhardt, P. J. \& Chaikin, P. M. Nature 436, 993Ð996 (2005).
\item Mikhael, J., Roth, J., Helden, L. \& Bechinger, C. Nature 454, 501--504 (2008).
\item Roichman, Y. \& Grier, D. Optics Express 13, 5434Ð5439 (2005).
\item Steinhardt, P. J. Nature 452, 43Ð44 (2008).
\item Ledieu, J. et al. Phys. Rev. B 72, 35420 (2005).
\item Glotzer, S.C. \& Solomon, M.J. Nature Mater. 6, 557-562 (2007).
\item Ueda, K., Dotera, T. \& Gemma, T. Phys. Rev. B 75, 195122 (2007).
\item Keys, A. S. \& Glotzer, S. C. Phys. Rev. Lett. 99, 235503 (2007).
\end{enumerate}

\end{document}